\documentclass[
preprint,
showpacs,preprintnumbers,
 amsmath,amssymb,
 aps,
prl,
floatfix,
]{revtex4-1}

\usepackage{graphicx}
\usepackage{dcolumn}
\usepackage{bm}

\begin{document}


\title{Coupling of Josephson Currents in Quantum Hall Bilayers}
%
%

\author{X. Huang}
\author{W. Dietsche}
\author{M. Hauser}
\author{K. von Klitzing}
\affiliation{ Max-Planck-Institut f\"{u}r Festk\"{o}rperforschung,
Heisenbergstra{\ss}e
1, 70569 Stuttgart, Germany}

\date{\today}

\begin{abstract}
We study ring shaped (Corbino) devices made of bilayer two-dimensional electron gases 
in the total filling factor one quantized Hall phase which is considered to be a coherent BCS-like state of interlayer excitons. Identical Josephson currents are observed at the two edges while only a negligible conductance between them is found.
The maximum Josephson current observed at either edge can
be controlled by passing a second interlayer Josephson current at the other edge. Due to the large
electric resistance between the two edges, the interaction between them can only be mediated
by the neutral interlayer excitonic groundstate.
\end{abstract}


\maketitle


Reducing the dimension of an electronic charge system from three to two leads to the emergence of intriguing phenomena like the integer and fractional quantum Hall effect  (QHE) \cite{Klitzing1980,Tsui1982} or special forms of superconductivity  \cite{reyren2007}.
Two-dimensional electron gases (2DEG) in GaAs quantum wells embedded in AlGaAs buffer layers are unique because of their extreme high purity  characterized  by electron and hole mobilities of several million $cm^{2}/Vsec$ \cite{Pfeiffer2003}. In recent years it has been demonstrated, both experimentally and theoretically, that bilayers consisting of two closely spaced 2DEGs in such a system can even show a condensation into a BCS state where (quasi)-excitons of full and vacant electron states play the role of the Cooper pairs \cite{Eis2004}. This novel state exhibits signatures of both the QHE \cite{Eisenstein1992}
and a Josephson-like coherent coupling between the layers \cite{Spi2000,Tie2008}. 

In this Letter we use contact pairs at two edges of a ring-shaped (Corbino-) device where, in the condensate state, the edges are isolated from each other for charge but not for exciton flow \cite{Tiemann2008a,Tie2008,Finck2011}.  We find that the critical Josephson currents measured at the two edges are identical although the edge lengths differ by almost a factor of three. If we apply Josephson currents to both edges simultaneously we find that the critical Josephson current measured at one edge can be more than doubled if the  current at the other edge has opposite polarity. This coupling of the Josephson currents is possible because the excitons (Cooper-pairs) in a bilayer system may be injected with different polarities which distinguishes it from  conventional superconductors. 


\begin{figure}
 \includegraphics[width=0.4\textwidth]{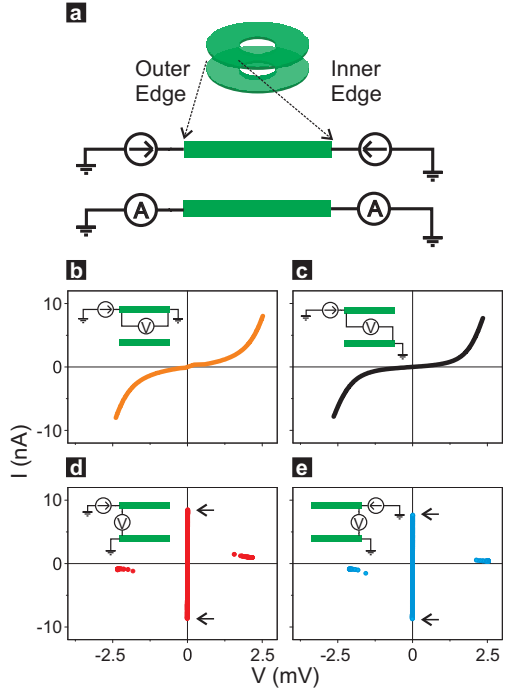}\\
 \textbf{\caption{\label{AllTunneling}
 {(a) Schematic of the Corbino device with a cross section. Variable current sources (arrows) are connected to the two edges at the top layers.  Voltages are measured using separate contacts (not shown).
 In (b)-(e), current-voltage characteristics are shown which are measured  in the $\nu_{T}=1$ state at temperatures below 50 mK. The respective measurement configurations are shown in the inserts. In the configurations (b) and (c), the conductances at small voltages are small. With contacts at the same edges (d) and (e), a Josephson-like I-V is measured indicating the exciton condensate. Note that the critical Josephson currents at the two edges are equal. }}}
\end{figure}

We study the excitonic condensates in bilayer quantum Hall systems consisting of two 2DEGs residing in 19 nm wide GaAs quantum wells separated by a 10 nm AlGaAs-barrier. Corbino rings with outer and inner diameters of 0.86 mm and 0.32 mm, respectively, are patterned by lithography. Front and back gates allow to adjust the densities of the two layers separately and to achieve separate contacts to the two layers.
Charge densities $n$ and the perpendicular magnetic field $B$ are set such that the filling fraction of the Landau levels $\nu=hn/eB$ is  approximately $1/2$  in each layer. At low temperatures (below a few 100 mK)  the electrons and holes (the occupied and the vacant states in the respective lowest Landau levels) become correlated and form excitons which condense in a BCS-like state \cite{Fer1989}, equivalently described by the ordering of pseudospins \cite{Yan1994}.

This state  prevails even if the tunneling probability between the layers is negligibly small as long as the $d/l_B$ ratio is less than about 2 where $d$ is the average distance between the 2DEG layers and $l_B={\sqrt{\hbar/eB}}$ is the magnetic length. Decreasing the  $d/l_B$ ratio enhances the coupling strength between the layers.
The correlated state is usually called the $\nu_{T}=1$ state 
because it is signaled by a quantum Hall effect with $\nu_{T}=1$ if standard Hall bars are studied  \cite{Kellogg2002,Tut2004,Wie2004,Yoon2010}.

We perform transport and tunneling measurements with Corbino devices schematically shown in Fig.~\ref{AllTunneling}(a).  Current-voltage (I-V)  characteristics are shown in Fig.~\ref{AllTunneling}(b)-(e) and are measured with the different contact configurations indicated by the respective inserts. In Fig.~\ref{AllTunneling}(b), current is injected into the outer edge of the upper layer and led out from the inner edge of the same layer. At small voltages, the electric conductance across the layer is only about $10^{-6}  S$ which is in agreement with earlier Corbino measurements  \cite{Tiemann2008a,Finck2011} in which only a very small charge current flows between the two edges. At voltages exceeding $\approx 1 mV$ the current starts to increase rapidly,   reminiscent of the breakdown behaviour of the traditional quantum Hall states. In the configuration Fig.~\ref{AllTunneling}(c) the I-V characteristic is qualitatively similar, mainly because it is dominated by the same large resistance to the flow of charges from edge to edge. 
 
 In contrast, the conductance increases by more than three orders of magnitude to $10^{-3} S$ if current contacts at the same edge are used. In Fig.~\ref{AllTunneling}(d) and (e) a Josephson-like tunneling I-V is observed which terminates at a critical current. Such behaviour is similar to the one observed earlier \cite{Spi2000,Tie2008}. The Josephson tunneling is a direct consequence of the interlayer coherence of the excitonic condensate \cite{Wen1993,Ezawa1994,Par2006}. This coherence develops in the bilayer systems after the layers become adjacent, in contrast to  "traditional" Josephson junctions where each of the two superconductors is already in a condensed state. In either case a coherence across a thin barrier exists which causes the dramatic enhancement of the tunneling conductance and allows a current flow across the barrier with no or very little dissipation, i.e. without a sizeable voltage drop across the barrier.
 
 It is important to note that both the critical currents measured at the two edges and the Josephson conductances in Fig.~\ref{AllTunneling}(d) and (e) are almost the same. Taking into account that the edge lengths differ by a factor of 2.7, this is evidence that the Josephson tunneling in this system is not an edge but a bulk phenomenon.  Moreover, the critical currents in Fig.~\ref{AllTunneling}(d) and (e) are not a bulk breakdown effect either, since that would require a ten times larger current. This was tested with samples in Hall-bar geometry from the same wafer and will be reported elsewhere \cite{Xuting2012}.

 The Josephson tunneling in bilayers can be viewed as an Andreev-like process \cite{Jun2008} where electrons injected into one layer "reflect" electrons out of the other layer, thereby forming additional excitons, i.e. extra electrons in one layer bound to extra holes in the other. The application of Josephson current is thus equivalent to the injection of excess bilayer excitons. Such excitons can distribute themselves in the sample bulk in form of a coherent counterflow current, which is sunk by the interlayer tunneling \cite{Su2010,Hyart2011}. While the excess excitons can recombine via tunneling, it might also be possible to compensate them by injecting oppositely polarized ones into the system. Corbino devices are unique for such an approach because charge currents and exciton flows are easily separated. 

 \begin{figure}
 \includegraphics[width=0.4\textwidth]{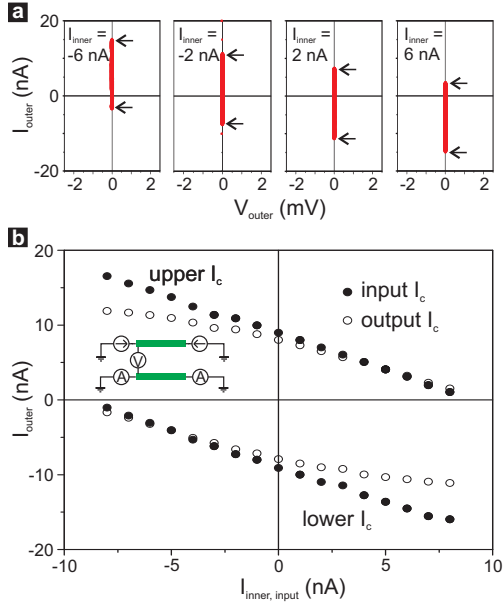}\\
  \textbf{\caption{\label{InnerCurrents} 
{(a) Josephson effect measured at the outer edge with different constant currents passing through the inner edge contacts. The critical Josephson currents through the outer contacts are increased (decreased) if an additional current flows through the inner contacts in the opposite (same) direction. 
(b) Upper and lower (positive and negative) input critical currents (black dots) measured at the outer edge as function of the Josephson current passing through the inner ones. The white dots are the critical currents measured at the output of the lower layer. They deviate from the input ones at large currents because large interedge voltages cause a parasitic charge current.}}}
\end{figure}

\begin{figure}
 \includegraphics[width=0.4\textwidth]{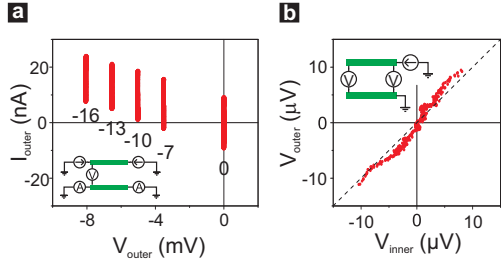}\\
  \textbf{\caption{\label{MaxCritCurrentPRL} {(a) Outer-edge Josephson I-V characteristics with a larger range of compensating inner currents. The I-V characteristics are offset by ($I_{inner}/2nA$) mV, respectively. The numbers at the traces are the respective $I_{inner}$ values (nA). The positive (upper) critical Josephson current can be more than doubled while the negative (lower) one is reduced and even changes sign with sufficiently large compensation. (b) Plotting the voltages at the outer edge vs. the one  at the inner edge while the Josephson current is increased. The two voltages are effectively identical.}}}
\end{figure}

We have realized this by using all four current contacts shown in Fig.~\ref{AllTunneling} (a). In the following data set the I-V-characteristic at the outer edge is measured as in Fig.~\ref{AllTunneling} (d) while a constant Josephson current $I_{inner}$ is passed across the two layers via two contacts at the inner edge. Results for different values of $I_{inner}$ are shown in Fig.~\ref{InnerCurrents}(a). Clearly, the tunneling I-V curves are shifted up and down depending on magnitude and direction of $I_{inner}$. 
Note that a negative (positive) current $I_{inner}$  increases (decreases) the respective positive (negative) critical current  $I_{c}$ at the outer edge. Note also that the respective change of the $I_{c}$ is equal to the Josephson current $I_{inner}$ applied to  the inner edge. In Fig.~\ref{InnerCurrents}(b)  the positive and negative critical Josephson currents are plotted (black dots) as function of the inner Josephson current. A linear dependence of the outer critical current on the inner current is observed.

As shown above, the two edges are separated for charge flow. Our device thus effectively possesses two separated Josephson contact pairs that are, however, coupled to each other. This is explained by the exchange of excess excitons generated by the Andreev process between the edges: applying a negative inner Josephson current is equivalent to injecting excess excitons which are oppositely polarized to those injected at the outer edge. Since the excitons share the same groundstate which extends all over the sample area a compensating process occurs leading to an increase of the critical current measured at the outer edge. In contrast, a positive inner current decreases the critical current at the outer edge, since the excitons injected at both edges have now the same polarity. Hence, beyond demonstrating that the "Cooper-pairs" which form the $\nu_{T}=1$ BCS state are polarized, our result provides possibly the first case in which two electric currents, while being separated from each other by a non-conducting bulk, are intricately coupled via the exchange of neutral objects.
 
In Fig.~\ref{MaxCritCurrentPRL}, the range of the compensating currents $I_{inner}$ is extended even further and exceeds the system's intrinsic critical current. This requires a positive current at the outer edge prior to the application of the inner current. Several I-V characteristics are shown measured at the outer edge with compensating currents down to -16nA (the voltages of the different curves are offset from each other). 
Amazingly, with compensating inner currents below -9nA, both ends of the Josephson current are now positive and the maximal critical current is more than doubled compared to the one at $I_{inner}=0$.
The compensation phenomenon is lost after increasing the compensating current beyond  $\pm16 nA$. Josephson tunneling is still observed but the I-V traces are symmetric as if no compensating current is applied.
It is intriguing to claim that at this maximal compensating current, the groundstate excitons flowing between the edges reach a critical velocity. This velocity would be 20 $cm/s$ if one 
takes the charge density in the layers as the density of the excitons and their flux as the one of the compensating electric current. This is much less than the pseudospin wave velocity ( $\approx$ 13 km/s) which would be the ultimate limit \cite{Hyart2011}.
Hence, it is more likely that a more trivial phenomenon limits the Josephson tunneling at large compensating currents, namely parasitic voltages which build up between the edges. 

The origin of these voltages are the unavoidable resistances between the bilayers and the contacts which lead to different voltage drops between the two different edges of the lower layer and the ground. 
These resistances originate from both the actual contacts and of the 2DEGs which leads to the bilayer device and are of the order of  $50 k\Omega$.
With  Josephson currents of $16 nA$ flowing in $opposite$ directions at the two edges one expects the build-up of an interedge voltage of $1.6 mV$. In this voltage regime the insulating properties of the bulk of the sample break down as one can see from Fig.~\ref{AllTunneling} (b) and (c). Thus, the two Josephson circuits can no longer be expected to remain electrically isolated from each other. This effect becomes already visible at smaller compensating currents.
In Fig.~\ref{InnerCurrents}(b), the white dots show the output critical Josephson currents measured at the lower layer of the outer edge. They coincide with the input ones (black dots) over a wide range of currents demonstrating that the Josephson currents at the two edges remain largely uncoupled as they should in the excitonic-condensate picture.
Deviations occur as soon as the difference between the Josephson currents at the two edges become large and a small charge current begins to flow between the edges.

In contrast to the interedge voltages, the interlayer voltages remain very small in the Josephson regime and are typically a few ${\mu}V$. This is demonstrated in Figure~\ref{MaxCritCurrentPRL} (b) which shows the outer edge voltage as function of the voltage at the inner edge while the Josephson current at the inner edge is varied. Strikingly, the two voltages are virtually identical although one would naively expect much smaller voltages at the outer edge than at the inner one. The apparent tunnel resistance at both edges is about 1 $k\Omega$ while the layer resistances to charge flow has been measured to be $\sim1M\Omega$. According to  Kirchhoff's law, the voltage at the outer edge should be reduced by a factor of 1000 compared to the one at the inner edge. The voltmeter at the outer edge is, however, connected to the inner one by the exciton condensate supporting a dissipationless counterflow of the charges which equilibrates the interlayer voltage.

In conclusion, we use the blocking of the charge current in the bulk of Corbino devices to study the interplay between the Josephson currents at the two edges. We find equal critical currents at the two edges which differ significantly in length indicating that Josephson tunneling extends over the whole sample area in agreement with earlier experimental and theoretical studies \cite{Fin2008,Tie2009,Eastham2010,Hyart2011}. This extension over the whole area leads also to the equal interlayer voltages at the two edges while the interlayer current is applied at one edge.

Applying currents to the inner and the outer edge of the Corbino device simultaneously, we find that they connect via the Andreev reflection to the same excitonic groundstate and the device appears as one Josephson contact although no charge current flows between the two current circuits (apart from parasitic effects). There is, however, a flow of coherent excitons between the edges which are polarized by the Andreev reflection process. Therefore, their density can be compensated or enhanced by a second circuit. 
As a consequence, the critical Josephson current measured at one edge can be more than doubled by injecting oppositely polarized excitons or reduced to less than zero, i.e. made to change sign,
 by equally polarized ones. Tunneling remains active as the intrinsic relaxation process of the excess excitons and determines the range between the upper and the lower values of the critical currents.
 
In principle, the coupling between the edges should also be observable by the ground-state phase coherence. However,  impurity-induced merons (vortices) probably cause a large number of phase jumps \cite{Hyart2011} making direct phase-sensitive measurements difficult. This is possibly the same reason why the expected Fraunhofer pattern of the Josephson current as a function of a parallel magnetic field \cite{Ezawa1994} remains elusive.

 We acknowledge discussions  with 
 A. H. MacDonald, 
T. Hyart,
B. Rosenow,
S. Schmult,
J. Smet,
and D. Zhang.
M. Hagel assisted in the sample preparation.
The experimental techniques are based upon earlier collaborations with J.G.S. Lok, L.Tiemann and W. Wegscheider. 
This project was supported by the BMBF (German Ministry of Education and Research) Grant No.s 01BM456 and 01BM900.

\end{document}